# A Point-process Response Model for Spike Trains from Single Neurons in Neural Circuits under Optogenetic Stimulation

X. Luo[a]*, S. Gee[b], V. Sohal[b], D. Small[c]

Optogenetics is a new tool to study neuronal circuits that have been genetically modified to allow stimulation by flashes of light. We study recordings from single neurons within neural circuits under optogenetic stimulation. The data from these experiments present a statistical challenge of modeling a high frequency point process (neuronal spikes) while the input is another high frequency point process (light flashes). We further develop a generalized linear model approach to model the relationships between two point processes, employing additive point-process response functions. The resulting model, Point-process Responses for Optogenetics (PRO), provides explicit nonlinear transformations to link the input point process with the output one. Such response functions may provide important and interpretable scientific insights into the properties of the biophysical process that governs neural spiking in response to optogenetic stimulation. We validate and compare the PRO model using a real dataset and simulations, and our model yields a superior area-under-the-curve value as high as 93% for predicting every future spike. For our experiment on the recurrent layer V circuit in the prefrontal cortex, the PRO model provides evidence that neurons integrate their inputs in a sophisticated manner. Another use of the model is that it enables understanding how neural circuits are altered under various disease conditions and/or experimental conditions by comparing the PRO parameters. Copyright © 2014 John Wiley & Sons, Ltd.

[a] *Department of Biostatistics, Brown University, Providence, Rhode Island 02912, USA*
[b] *Department of Psychiatry and Neuroscience Graduate Program, University of California, San Francisco, California 94143, USA*
[c] *Department of Statistics, The Wharton School, University of Pennsylvania, Philadelphia, Pennsylvania 19104, USA*
* *Correspondence to: Department of Biostatistics, Brown University, Providence, Rhode Island 02912, USA. E-mail: xi.rossi.luo@gmail.com*

*Contract/grant sponsor:* The National Institutes of Health [P01AA019072, P20GM103645, P30AI042853, R01NS052470, S10OD016366]; Brown University [seed award]; and Brown Institute for Brain Science [pilot award].

## 1. Introduction

In 1979, Francis Crick suggested it would be of great value for understanding how the brain works to be able to deliver an input to a specific neuron in the brain at a specific time while leaving other neurons unaltered, and then to observe the downstream effects of the input [1]. Optogenetics is a new technology for accomplishing this goal. It works by using genetic tools to make neurons sensitive to light, and then using light flashes to deliver inputs in a fast and targeted way [2]. One application of optogenetics is to study how neural circuits (networks) process their inputs. A neural circuit is an ensemble of interconnected neurons that work together to process specific kinds of information. For circuits involved in early stages of sensory processing such as those in the retina, the way in which the circuits process information was studied before the invention of optogenetics by delivering controlled sensory stimuli and using statistical models to infer the input features that the circuit's output (i.e., the spike trains of the neurons) encodes [3, 4]. However, for circuits that are deeper in the brain (e.g., those in the prefrontal cortex involved in working memory, attention, cognitive control, and social cognition), precise control of the inputs to the circuits has only been made possible recently by the development of optogenetics. We will consider a specific experiment to understand the processing of information in the recurrent layer V circuit in the prefrontal cortex [5]. Malfunctions of this circuit are thought to play a role in schizophrenia, a devastating illness that affects approximately 1% of the population worldwide [6, 7, 8].

In our experiment, a brain slice that contains a recurrent layer V circuit from a mouse is used, light flashes are delivered to the circuit of infected neurons at randomly chosen times, and the spikes of a non-infected neuron are recorded. Neurons that are directly activated by light may recruit other excitatory and inhibitory neurons in the circuit, producing a combination of excitatory and inhibitory input that continues to reverberate long after the original stimulus ends. A raster plot of the dataset is shown in Figure 1. Each line shows the stimuli (light flashes) and outputs (neuronal spikes) for one run (sweep) of the experiment. The recorded neuron within a network shows complex patterns of firing that may appear to be "noisy". This firing behavior is different from a typical isolated neuron, which, for example, produces single spikes responding to light flashes with a constant latency of about 10ms and almost no extraneous spikes [9]. The scientific questions we seek to answer are the following: (1) what roles do the presence or absence of recent stimulation (light flashes), the past history of stimulation and the past neuronal activity (i.e., the times of previous spikes) play in the probability of spiking at a given time; (2) how predictable are the spikes based on the light flashes, i.e., how much noise is there in how the neurons process their inputs; and (3) how do (1) and (2) differ between different types of neurons in the circuit (neurons in the circuit can be subdivided into different types based on morphology and electrophysiological properties) or when different drugs are applied to the circuit. In this paper, we further develop a statistical model for spike train data to address the first two questions, and make an initial attempt to address question (3) via examining the impact of varying biophysical characteristics of simulated neurons on the fitted parameters of our model.

In this paper, we focus on building a model for the firing pattern of a single neuron under network stimulation. Understanding the firing pattern of a single neuron under network stimulation is a useful step in studying and modeling the behavior of whole neural circuits [5, 10]. Such understanding is especially important for optogenetic stimulation due to its control at the single-neuron, single-spike level [11]. Our model for the firing pattern of a single neuron under network stimulation builds on existing spike train models that were developed for other types of experiments but incorporates new elements that are motivated by the specific features of optogenetics stimulation of neural circuits. Indeed, probabilistic approaches have been successfully applied to model neuronal data. Generalized linear models (GLM) have been used to link the input stimuli with the output spikes [12, 13, 14, 15, 16, 17], see [18, 19] for review. The spike process can be modeled by Poisson [20] for spike counts per bin or Bernoulli [14] for binary sequences. Then linear combinations of basis functions are used to approximate the delayed effects of the past spikes and stimuli, see Section 3.3 for a more detailed description. These approaches are flexible in general because a large number of bases can be used. For randomly generated stimuli and sparse spikes, the applicability of these methods can be challenging due to limited data available for estimation. Bayesian approaches using regularization [19] can help circumvent this issue, but rely on proper choices of bases and regularization terms, in order to avoid large biases introduced by regularization [21]. We instead design three interpretable basis functions, and these functions are especially useful for situations where both the stimuli and spike sequences are sparse, see Figure 1 and more description of the data in Section 2.

There are a few scientific reasons for studying such sparse data. Optogenetics for neural circuits is an emerging technology, and a fixed input pattern that is the most scientifically meaningful has not been established yet, partly because circuits may have complex responses behavior than single neurons. Instead, one objective is to understand how neurons respond to different temporal patterns of input, in particular, inputs which activate many neurons in the local circuit, eliciting high levels of both inhibitory and excitatory currents, similar to the situation in vivo. Moreover, some investigators are interested in inferring the overall effects of experimental conditions (e.g. medication) that can be generalized to study various inputs, rather than restricting the conclusions based on a fixed input pattern [5]. For random stimuli, the GLM approaches for binary inputs (e.g. [14]) may be applied but will suffer from the sparse data issue as discussed before. Alternatively, the spike-triggered average statistic [22, 23] has also been proposed before for random Gaussian inputs. However, this statistic is usually used as the first step in analysis due to its limited on-line interpretation and response variability, and the applicability to non-Gaussian stimuli is not well understood [22].

Along with the sparsity in the data, the probability (or after transformation, e.g. logit) of spiking as a function of time elapsed since the last stimulus can be highly nonlinear [12, 4, 24]. In the statistical literature, nonparametric approaches have been applied to account for such nonlinear underlying processes. Brillinger [12] considers higher order polynomials in fitting the spiking probability. A recent review by Kass and colleagues [24] suggests using splines. In the existing GLM approaches, other basis functions have been employed, see [18, 19] for review. From a purely methodological point of view, estimating nonlinear functions using multiple basis functions from sparse design points is particularly challenging. Seifert and Gasser [25] show that one needs sophisticated procedures to overcome such challenges in theory and practice, when the design points are sparse. Another remedy is to add additional points [26]. However, this remedy is not available in our experiment, as the input processes are randomly generated and the output processes are not controlled by the experimenters. As a consequence, the dataset yields mostly non-repetitive flash patterns as predictors. For example, only a small number of samples with the same or similar flash patterns are observed, and the observed patterns cover only a small portion of all possible patterns that could arise. In this paper, we adopt a different and simple approach to model the outcome instead. We construct three simple non-linear functions, called point-process response functions, to summarize the important statistics of these flash patterns. These functions will be shown to provide good fit and yield important biophysical interpretations.

Response functions have been previously applied for analyzing and modeling classical spike train data, also known as filter functions, receptive fields, or response kernels. Gerstner, Kistler, and colleagues [27, 28, 29] introduce the Spike Response Model (SRM) based on the Integrate-and-Fire (IF) model [30]. By integrating the IF model, they replace the integrands by a general response kernel function, independent of the input. Keat, Geffen, and colleagues [4, 31] also construct a linear-nonlinear model for spike train data, and then a generating potential is constructed by convolving an impulse response function with the stimulus process. Paninski, Pillow, and colleagues [32, 33] propose a stochastic IF model using kernel projection functions, and estimate the kernel functions using maximum likelihood. Pillow and colleagues [15] study raised cosine basis functions using the GLM framework for spike train data. The applicability of these latter two approaches will be compared using the experiment data and simulated data.

In summary, in the setting of such sparse responses and inputs, we introduce Point-process Responses for Optogenetics (PRO) to parametrically model the spiking probability. This approach has several nice features. The simple form avoids overfitting, yet provides highly generalizable models. Moreover, this approach also provides additional advantages in interpretation from the fitted coefficients. In addition, the computation is very fast using existing logistic regression implementations. Finally and perhaps most importantly, the outputs from this model can be related to other biophysical models, and the model coefficients reveal a few important aspects of how the circuits processes inputs. Furthermore, the ways in which different types of neurons process information or how a drug affects information processing can be assessed by comparing the coefficients between different neurons or under different drug treatments.

Our approach also provides highly accurate prediction of future spikes. Indeed, several papers have measured prediction accuracy of various neuronal models, see [4] and [33]. Jolivet and colleagues [34] study the prediction performance of the nonlinear IF model and the SRM, and typical prediction accuracy is 70-80%. We will apply an out-of-sample evaluation approach, where we aim to predict what will happen in the future given the history up to now. Using this benchmark, the PRO method gives high area-under-the-curve (AUC) values for real data, in the high 90%.

The rest of the paper is organized as follows. We describe the dataset in Section 2. We then motivate the forms of the response functions in Section 3. In Section 4, we compare the PRO approach with other methods using an optogenetics experiment, and we also consider variation of the parameters in our approach. All these approaches are further compared using simulated data from a classic neuronal model and a simple network model in Section 5, and the biophysical interpretations of the PRO models are also illustrated. Finally, we discuss future directions in Section 6.

## 2. An Optogenetics Experiment

The optogenetics experiment protocol has been previously described in [35] and [5]. The input stimuli consist of trains of short flashes of blue light (470nm). The presence of a flash within each 5ms interval is coded by a 1 if a flash is present and 0 otherwise. The timing of such flashes are randomly generated from independent Bernoulli random variables for each 5ms interval. Spiking is also binary coded for each 5ms interval, where 1 or 0 indicates the presence or absence of a spike, respectively. Each 5ms interval contains at most 1 flash or 1 spike. Thus we obtain two binary sequences of spikes and flashes for each run.

We use 5ms bins because this is a reasonable time resolution to model the behaviors of neurons which rarely spike at rates > 100 Hz. We choose to model spikes instead of membrane potentials, because this simplifies the analysis and spikes represent the relevant output from neurons.

Figure 1A illustrates the raster plot of the spikes and light flashes in an experiment in which 20 light trains are delivered on consecutive runs ("sweeps"). We observe 169 spikes in the first 10 sweeps, and 181 spikes in the next 10 sweeps. This is a result of 686 flashes in the first 10 sweeps and 691 flashes in the next 10 sweeps. The total number of time points (5ms bin) for each sweep is 500. There is also some variability of spike counts for each sweep, ranging from 12 to 24 with a median of 17.

It should also be noted that the input sequence for each sweep is different as they are randomly generated, and overall the input stimulus sequence has low repeatability. After omitting the first spike in each run, we have a total of 330 inter-spike segments out of 350 total spikes over 20 sweeps. 321 inter-spike flash patterns are non-repeated, 8 patterns are repeated once, and only 1 pattern is repeated twice. In Figure 1B, the inter-spike times, regardless of flash patterns, also show some variation as well, and most of the inter-spike times are repeated less than 5 times. Some of them are larger than 250 ms, but are repeatedly observed no more than two times. The median interspike interval is 115 ms, which is approximately an order of magnitude longer than the typical membrane time constant for pyramidal neurons. In Figure 1C, the inter-spike flash counts also vary from 1 to 9, with a median of 4.

In all of the following analyses, we will first divide the data into two parts to test model generalizability. The first 10 sweeps will be used as a training dataset to fit models, and the last 10 sweeps will be used as a testing dataset to validate the models.

## 3. Model

We will first introduce some basic assumptions in modeling these point processes under a probabilistic framework. We will then motivate the parametric forms of our response functions, and relate these to their biophysical interpretations. A spline nonparametric approach will also be introduced to assess the residual nonlinearity effect not captured by these response functions.

### 3.1. Model Assumptions

To model the effect of the input sequence, we start with the following general probability model

$$P\left(s_t = 1 | \mathbf{s}_0^{t-1}, \mathbf{f}_0^t\right) = H_t\left(\mathbf{s}_0^{t-1}, \mathbf{f}_0^t\right) \qquad (1)$$

where $s_t$ is the spiking status at time interval $t$ (1 means spiking, 0 otherwise), $\mathbf{s}_0^{t-1}$ is the sequence of spiking statuses from time interval 0 to time interval $t-1$ (i.e. $\mathbf{s}_0^{t-1} = (s_0, s_1, \ldots, s_{t-1})$), and similarly $\mathbf{f}_0^t = (f_0, f_1, \ldots, f_t)$ with $f_t$ denoting the light stimulation status at time interval $t$ (1 means light flashing, 0 otherwise). Without loss of generality, $t$ is assumed to be nonnegative integers. Note that we may include $f_t$ as a predictor for the response $s_t$, because $s_t$ is usually believed to happen immediately after, responding to $f_t$, even if they fall on the same time interval.

Model (1) in its general form assumes no reproducibility of spiking. It is less useful in practice because it does not provide information about the mechanisms of neuronal spiking nor does it provide predictive systems for future spikes. We make two important assumptions to simplify the forms of $H_t$, such that this function can be easily estimated from the data. These two assumptions have been employed in modeling spike train data, see [20] for example.

The first assumption is that the function $H_t$ depends only on the time $t^*$ of the last spike and the flashes $f_t$ back to the last flash time $t^\ddagger$ before $t^*$. That is

$$H_t\left(\mathbf{s}_0^{t-1}, \mathbf{f}_0^t\right) = H_t\left(t^*, \mathbf{f}_{t^\ddagger}^t\right). \qquad (2)$$

This assumption has often been used [36, 20, 37, 29, 38], and the resulting process is called a renewal process sometimes. This assumption roughly coincides with our scientific hypothesis that the neurons reset the membrane potential to zero after every spike, but we here allow this dependence to be slightly longer back to $t^\ddagger$, because it may contain information on how the sparse inputs are integrated in neurons. This is also due to the hypothesis that the sensitivity of neuronal integration may depend on the spacing of flash times, which is independent of the membrane potential reset. In practice, when we estimate $H_t$ from the data, this assumption fits our data slightly better, see Section 4.4 for a empirical comparison.

A special case of Equation (2) is that it takes an additive form after a logit transformation, which is our second assumption. That is, we assume

$$\text{logit}\left(H_t\left(t^*, \mathbf{f}_{t^\ddagger}^t\right)\right) = \beta_0 + \sum_{k=1} \beta_k h_t^{(k)}\left(t^*, \mathbf{f}_{t^\ddagger}^t\right) \qquad (3)$$

where $h_t^{(k)}\left(t^*, \mathbf{f}_{t^\ddagger}^t\right)$ is a response function that models the input effects, $\beta_k$ is its coefficient, and $\beta_0$ is the intercept. We will construct these response functions parametrically in Section 3.2, and a nonparametric extension will also be considered.

The assumption of additive forms simplifies the computation, and it is related to generalized additive models (GAM) [39]. Moreover, the logit transform is approximated by the logarithm transform if $H_t$ is small, and thus the formation (3) mimics a similar additive form of a special case, multiplicative inhomogeneous markov interval process [20]. Additive forms have also been employed before [12, 13, 14], see [18, 19] for review.

Intuitively, this logit transform can be understood to model sharp rises in spiking probability, when for example the right hand side of Equation (3) could be exponentiated in order to approximate the probability scale. The sharp rising phenomenon is considered appropriate for neuronal data under a few biophysical models. For example, the linear-nonlinear model employs a sharp rising non-linear function, similar to the shape of the logistic function, to transform the linear components related to the stimuli. Finally, a critical step of fitting the likelihood under Equation (3) is to provide or estimate continuous function atoms $h_t^{(k)}$ from the flash point process, and we turn to it now.

### 3.2. Choice of Response Functions

We propose a simple formulation of the summation in (3), which contains the following 3 components

$$h_t^{(1)} = h_1\left(t - t^\dagger\right), \quad h_t^{(2)} = h_2\left(\sum_{\tau=t^*}^{t} f_\tau\right), \quad h_t^{(3)} = h_3\left(\mathbf{f}_{t^\ddagger}^t\right) \tag{4}$$

where $t^\dagger$ is the last flash time. These functions will be constructed parametrically or non-parametrically. The first component describes the time elapsed since the last flash. The second component characterizes the total number of flashes since the last spike. The third function measures the spacing of flashes since $t^\ddagger$.

As described before, these three functions can be estimated from the data using general GAM software. We prefer to construct these functions explicitly in parametric forms, because they are computed faster, provide better interpretability, and can be embedded in other statistical procedures for inference. For example, statistical testing for the coefficient changes in (3) under different conditions would yield important insight on how the behavior of neural circuits is altered. As a complimentary procedure and validity check, we will also compare with the nonparametric estimates.

We now introduce the parametric forms of the functions in Equation (4). Let $t_{(j)}$'s, for $j \geq 1$, be reversely ordered previous flash times $F_t = \left\{\tau : f_\tau = 1, \tau = t^\ddagger, \ldots, t-1, t\right\}$ since $t^\ddagger$. We set $t_{(0)} = t$ for notation simplicity. The three functions are

$$h_1\left(t - t^\dagger\right) = \log\left(1 + t - t^\dagger\right),$$
$$h_2\left(\sum_{\tau=t^*}^{t} f_\tau\right) = \log\left(1 + \sum_{\tau=t^*}^{t} f_\tau\right),$$
$$h_3\left(\mathbf{f}_{t^\ddagger}^t\right) = \log\left[\log\left(1 + \sum_{j=1}^{|F_t|}\left(t_{(j-1)} - t_{(j)}\right)^2\right)\right].$$

To ensure existence of the logarithm transform, a small value 1 is added to the base quantities in the above. We take the logarithm transform and the log-log transform, because they spread out the clusters of small values and have interpretations of relative changes rather than absolute changes. Empirically, they fit the data well, and are validated using a GAM fit.

The interpretation of $h_1$ and $h_2$ is straightforward. They are logarithmically transformed values of the elapsed time since the last flash and the total number of flashes since the last spike respectively. We shall call the corresponding coefficients post-flash (PF) and cumulative-flash (CF) parameters. The third function $h_3$ is measuring the spacing of these flashes since $t^\ddagger$, conditional on $|F_t|$ flashes, $t^\dagger$, and $t^\ddagger$, due to Lemma 1. We shall call the corresponding coefficient spread-flash (SF) parameter. Figure 2 shows the schematics of these 3 functions, responding to an example sequence of flashes and spikes. Note that these 3 functions do not depend on the current spike status, $s(t)$, and thus they form a valid design matrix for predicting future spiking.

We have the following lemma concerning the extreme values of the SF function $h_3$.

**Lemma 1.** *Fix $|F_t| > 2$, $t^*$, $t^\ddagger = t_{(|F_t|)}$ and $t$, such that $t - t^* \geq |F_t| - 1$. The SF function $h_3$ achieves its maximum when (i) $t_{(1)} = t$ and $t_{(|F_t|-1)} = t - |F_t| + 2$ if $t^* > t^\ddagger$ or (ii) $t_{(1)} = t$ and one and only one $t_{(j-1)} - t_{(j)} = t - t^\ddagger - |F_t| + 2$ if $t^* = t^\ddagger$. The SF function achieves its minimum when $t_{(j)} - t_{(j-1)} = \lfloor (t - t^\ddagger) / |F_t| \rfloor$ if $j \in S$ and $t_{(j)} - t_{(j-1)} = \lceil (t - t^\ddagger) / |F_t| \rceil$ otherwise, where the set $S$ satisfies $|S| = |F_t| \lceil (t - t^\ddagger) / |F_t| \rceil - (t - t^\ddagger)$, provided that $t^* - t^\ddagger < \lfloor (t - t^\ddagger) / |F_t| \rfloor$.*

The implication of Lemma 1 is intuitive. Simply, this function is maximized at time $t$ if all the past flash times are as close to each other as possible, conditional on all other quantities are fixed in the lemma, including the number of flashes since the last spike being quantities that are fixed and nonzero, the last flash time before the last spike being fixed. Especially when $t^* > t^\ddagger$, the maximum is achieved when all flashes since the last spike occur close to time $t$. This term can be used to answer a fundamental question about how neural circuits integrate their inputs, namely, do they care about the patterns of input or does only the total amount of input matter? The coefficient on SF is positive, see Table 1 later, The scientific interpretation behind this is that the spiking probability is maximized when the neurons are continuously flashed after a period of no flashes since the last spike, if the period is longer than the refractory period. More explanations of this effect as well as its exceptions will be discussed later. Conversely, this function achieves its minimum if the spacing between the flash times are approximately as even as possible, when the cumulative stimulus effect from the number of flashes is washed out due to time decay.

Another intuitive understanding of SF $h_3\left(\mathbf{f}_{t^\ddagger}^t\right)$ with respect to $t$ may also be obtained as if the time $t$ is continuous. Take the derivative of $h_3\left(\mathbf{f}_{t^\ddagger}^t\right)$ with respect to $t$ to yield

$$\frac{\partial}{\partial t} h_3\left(\mathbf{f}_{t^\ddagger}^t\right) = \frac{t - t_{(1)}}{A_1 A_2} \qquad (5)$$

where $A_1 = 1 + \sum_{j=1}^{|F_t|} \left(t_{(j-1)} - t_{(j)}\right)^2$ and $A_2 = \log A_1$. The denominator is to discount the increment if $A_1$ (and thus $A_2$) is large. The numerator in Equation (5) shows that the increase is large if the previous flash $t_{(1)}$ is far from the current time $t$ and zero if $t = t_{(1)}$. When $t - t_{(1)}$ is large, the large increment is due to a long period of non-flashing, and also because the current value is far below the maximum as shown in Lemma 1. When $t = t_{(1)}$, it is a necessary condition for a (local) maximum, and the global maximum is proved in Lemma 1 with this being one of the conditions (and the other is a large non-flashing period to produce a large increment). For discrete time $t$, these can be regarded as approximations to the discrete increments.

Because SF measures the spacing of flashes conditioning on the number of flashes, it is then viable to consider including the interaction between CF and SF. Indeed, we found this interaction term is significant when fitting the experiment data, see Section 4. It is also possible to consider even higher order terms of these 3 predictors. For example, one may consider their polynomial interactions. We have examined such possibilities up to the third order, but found only the interaction term between CF and SF survive forward/backward model selection with the AIC criterion.

A variation of this parametric PRO model is to estimate these functions non-parametrically, for example using splines [24]. Given the history of generalizing parametric models to non-parametric or semi-parametric ones, we see that this approach is a straight forward extension of the parametric PRO model. We thus omit the detailed description on this extension here. Following [24], we choose the spline bases to fit GAM, and the generalized cross validation criterion is used to pick the smoothing parameter, as implemented in the R package *mgcv* [40]. Because PRO discovers a significant interaction between CF and SF functions, we will consider fitting a smoothing plane of CF and SF, in addition to a smoothing function of PF. This spline extension of PRO is named as PROs.

### 3.3. Comparison with Existing GLM Approaches

Generalized linear models for spike trains has been employed before for modeling spike train data, see [18, 19] for review. In a single neuron setting, an important class of model based on filters can be summarized as

$$\log \lambda_t = \mathbf{r}_k \cdot \mathbf{x}_t + \mathbf{r}_h \cdot \mathbf{y}_t + \mu \qquad (6)$$

where $\lambda_t$ is the conditional fitting rate at time $t$ conditional on the past history of spikes and stimuli, $\mathbf{r}_k$ is the stimulus filer, $\mathbf{x}_t$ is the stimulus history vector at time $t$, $\mathbf{r}_h$ is the post-spike filer, $\mathbf{y}_t$ is the spike history in a window just preceding time $t$, and $\mu$ is an intercept term characterizing the baseline fire rate. The firing rate $\lambda_t$ can be used in both Poisson distributions and Bernoilli distributions. In the Bernoilli case, $\lambda_t$ models the probability of spiking within a time bin $dt$ as $P(\text{no spike}) = \exp(-\lambda_t dt)$ [17]. Additional model parameters and link functions can be considered in this framework [18], but they are less applicable here for our experiment.

The filter functions $\mathbf{r}_k$ and $\mathbf{r}_h$ can be effectively modeled by a linear combination of multiple cosine basis functions [15, 16], and the estimation can be carried out by maximum likelihood for a class of cascade models [32]. Alternatively, these can be estimated parametrically with one coefficient for each time lag [14], which mathematically is equivalent to choosing the bases of Kronecker delta functions $\delta(t - \tau)$ for various time lag $\tau$, and the maximum lag can be selected by AIC. In general, all these approaches involves estimating several model parameters, one coefficient for each basis function. For example, 10 cosine bases are recommend for the stimulus and post-spike filters respectively as a general guideline [15, 17]. If the stimulus history $\mathbf{x}_t$ and the spike history $\mathbf{y}_t$ explore the design matrix well, and the collection of basis functions provides a close approximation to the underlying filters, one would expect these methods estimate the filters well. However, this may over fit for limited data. For example, in our experiment the flash patterns are non-repetitive and the spike patterns are sparse, see Figure 1. We expect this kind of sparsity to be common in many optogenetics experiments on neural circuits.

Our approach, in a certain aspect, can be regarded as designing some special bases that can be interpreted well for sparse data in our optogenetics experiment. It has only a minor deviation from Model (6) in that we use both $\mathbf{x}_t$ and $\mathbf{y}_t$ to determine the response function, which has been considered in renewal processes before [38]. Though the cosine basis in existing GLM approaches can be related to period or latency, this nonparametric basis to us has limited interpretation relating to the sparse patterns. In general their coefficients are rarely interpreted separately, and rather the resulting linear combination is a popular choice to compare different neurons or neurons under different conditions [18, 19]. In our model, these PRO coefficients can be compared, and we will illustrate this using simulated LIF neurons, see Figure 7.

In all the above approaches, the assumptions are that the filter model (6) captures the properties well and the basis functions are rich enough to model the filters if no interpretation of individual coefficients is required. For sparse data like ours, it would be difficult to check or test these assumptions statistically. As an approach for validating and comparing different models, we will employ cross validation on both real data and simulated data. Furthermore, it is also critical to avoid over-fitting in modeling sparse data. Bayesian approaches have been proposed to regularize the fit [18, 19] in the existing GLM methods, and we take the AIC regularization to select from higher order polynomial terms. Certainly, regularization on a general class of basis functions is a flexible approach in general, but its performance relies on the effective projection onto these bases (usually requiring the coefficients to be large enough, or the basis representation to be efficient) to overcome the bias introduced by regularization [21].

Due to our specially designed basis functions, we do not observe significant performance drop in our PRO method compared with our PROs method, and the residual nonlinearity as captured by PROs is not large (see Figure 4 for an example).

## 4. Analysis of an Optogenetics Experiment

In this section, we will employ PRO and PROs to analyze the neuronal spiking behavior under an optogenetics experiment. The generalizability of these two approaches will be compared with other methods. Especially, we will compare with two off-the-shelf predictive models, random forest (RF) and support vector machine (SVM), because they have become popular methods for computational neuroscience; a generalized linear model using filter bases (GLF) proposed by Pillow and colleagues [15]; and an important biophysical model, Generalized Integrate-and-Fire (GIF) model [32, 33], which extends the widely used IF model. We use the MATLAB implementations of GIF and GLF provided with the papers, R packages *e1071* (version 1.6) and *randomForest* (version 4.6-6) for SVM and RF respectively, and our own R implementation of PRO and PROs. We use the default parameters provided by these software programs. In particular, SVM uses radial kernels, and RF uses 500 trees. GIF and GLF use filter lengths of 150 ms and 100 ms under our sampling rate respectively, comparable to the median inter-spike time (115 ms) of our methods.

### 4.1. Model Fit

We first apply the PRO model to the first 10 sweeps of the optogenetics experiment illustrated in Figure 1. We then validate the fitted model using the next 10 sweeps. The prediction performance is compared with RF, SVM, GLF, and GIF.

We use the predictors PF, CF, and SF, which are based on only the history of the spike and flash time series, as explained in Section 3. The model fit from the first 10 runs is shown in Table 1. The actual values of these coefficients may be of less importance, because these values can change depending on the normalization of predictors. We suggest using this model to compare how the coefficients change under different experimental conditions, because this may yield insights into how various manipulations (e.g. drugs that block specific ion channels) affect the net input-output properties of the neuron. For example, if after drug application, the CF parameter decreases while PF remains constant, it would suggest that neurons are firing less because of a decreased ability to integrate multiple light flashes, rather than a decrease in the influence of a single light flash. We illustrate this kind of approach by fitting the PRO model to results of a simulation while varying the parameters in a leaky integrate and fire model in Section 5. This interaction model is selected using forward-backward variable selection with the AIC criterion, where we start from a full model with all possible third order interactions. The correlations between these 3 predictors are -0.24, 0.37, 0.47, where PF and SF has the highest correlation. The deviance R-square of this model is 36.1%. This shows that PF, CF, SF and the interaction between CF and SF are all significant predictors of spiking, as shown by the small p-values.

The large CF coefficient in Table 1 provides evidence that these neurons are integrating their input such that spiking depends on the pattern of flashes over time. Figure 3 plots the predicted probabilities and the 95% confidence intervals for both PRO and PROs for each time point in the example sequence of flashes and spikes in Figure 2. The integration of input is supported by the fact that the probability of spike rises with more flash inputs in general. The amount of increment, however, depends on the arrangement of flashes, as described jointly by the model coefficients CF and SF, and a flash closely trailing the previous one elicits a smaller increase than a flash far from the previous one. It is notable to point out that the period of integration in our experiment is longer than a typical membrane constant for neurons (around 20 ms). However a detailed investigation of the physical and chemical mechanisms that may contribute to this phenomenon is beyond the scope of this paper. In Section 4.2, we will assess if our model based on inter-spike intervals and flash patterns better predicts future spikes than a fixed window model with short or long windows. This finding that the cortical neurons being studied integrate their input is scientifically interesting and will be further discussed in Section 6.

As discussed before, instead of the parametric PRO model, the PROs model employs spline bases and GAM to estimate the response functions from the data. In Figure 3, the major difference between PRO and PROs fits is within the decaying period after each flash, especially immediately following a spike. This difference may be caused by a slight deviation from the 3 functions, as we will examine in detail later (Figure 4). However, this difference contributes little to predicting spikes, because it exists only within the non-spiking period and the predicted spiking probabilities are small. For the time bins immediately before a spike, the difference between PRO and PROs is almost non-negligible. Figure 4 further illustrate the difference between PRO and PROs by plotting both the partial regression function of PF and the fitted surface on CF and SF with fixed PF at its mean. The partial regression plot (Figure 4a) confirms that a linear function describes the contribution from PF well, except for a drop from the linear trend at the origin where PF=0. We think this could represent the delay for a flash to activate light-sensitive ion channels and thereby cause neural depolarization immediately after each flash. Notably, the fact that the coefficient for SF is positive suggests that closely spaced light flashes (conditional on the time past the last spike and other parameters as stated in Lemma 1) have a synergistic influence on neural excitability. The fitted surface plot (Figure 4b) confirms that the main terms of CF and SF capture the relationship well, except at the large ends of these two predictors. In that regime, the surface increases to a plateau when both are large, which could indicate that the stimulating contribution should be discounted if quite a number of flashes are generated very close to each other. One possible biophysical interpretation is that the circuit may not fully integrate such fast inputs, which may be a result of the inactivation of light-sensitive ion channels. This plateau observation is also consistent with the negative fitted coefficient of the interaction term in the parametric PRO model.

The spline extension PROs seems to be able to capture very minor deviations from PRO, as shown in Figure 3-4. However, including these does not improve prediction much, as we will see below. Because PROs provides less straight forward output for studying the properties of neurons, we recommend this approach only for scientists who are comfortable interpreting such spline models. PROs is also recommended for scientists who are interested in modeling those slight variations.

### 4.2. Model Validation and Prediction Comparison

To further validate the model, we use out-of-sample prediction as a criterion. We use the first 10 sweeps as the training data to build models, and test the model validity on the next 10 sweeps. We adopt the popular receiver operating characteristic (ROC) criterion as the measure of the prediction performance, because it provides a unified framework to evaluate various probabilistic and non-probabilistic methods to which we will compare the PRO method. If a model approximates the underlying firing mechanisms, we believe that the model have good prediction power for forecasting future spikes based on the past history. For comparison purposes, the same training/testing procedure is carried out for RF, SVM, GIF, and GLF. The same three response functions, PF, CF, and SF, are used as predictors in RF and SVM.

The predicted spiking probabilities of PRO, PROs, RF, SVM, and GLF are compared with the true spiking sequence. The GIF model provides instead the estimated membrane potentials, which will be used as surrogates of the spiking probability. Because the neural firing depends on the highest membrane potentials within each 5ms interval, the maximum estimated membrane potential will be used.

Figure 5 shows the ROC plot comparing these statistical and biophysical models. PROs outperform all other models using the area-under-the-curve (AUC) measure, followed by GLF and PRO. Their AUC values are 94.2%, 93.6%, and 92.7% respectively. Though these three models have comparable AUC values, PRO uses much less number of parameters. Our models PRO and PROs are much faster to compute as we demonstrate in Table 4. Moreover, nonlinear models such as PROs and GLF have only slightly higher AUC values, and the GLF AUCs are among the smallest in two simulated settings (Table 3). Therefore, we recommend the PRO model for scientifically-oriented users, because of its clear advantage in interpreting coefficient changes under different conditions, which may be difficult using other basis functions. GIF is the next best predictive model (AUC=88%), but its interpretation is again complicated by its complexity of modeling. For example, the GIF model involves estimating stimulus filter and spike-history filter functions among over a dozen parameters, see Section 4.2 for a detailed description. As a result, it is difficult to assess how the flash patterns drive the spike outputs by comparing these correlated estimates. For example, the question of whether neurons are integrating their inputs (i.e., whether closely or sparsely spaced spikes would encourage more spikes) could be hard to address using either the GIF or GLF model. On the other hand, the PRO coefficients provide simple and direct answers.

### 4.3. Expanding the Set of Predictors in PRO

To assess if the existing PRO model can be improved, we consider expanding the predictors included in the model. We consider with the following two scenarios: (1) adding the post-spike time to the model; (2) expanding to all third polynomials and interactions of the 3 functions. The resulting prediction performance is comparable to PRO in both scenarios, with AUC values 92.8% (95% CI: 91.4%-94.1%) and 92.8% (95% CI: 91.5%-94.2%) respectively. In both scenarios all the added predictors are not significant (p values > 0.56). This shows that the prediction performance of PRO cannot be increased significantly by adding these simple predictors.

### 4.4. Variation of History Dependence

Our models use the history information back to the previous spike time $t^*$ and the previous flash time $t^\ddagger$ before the previous spike. When we use $t^*$ in place of $t^\ddagger$ when defining our response functions as in [24], the residual deviance of our PRO model increases from 895.38 to 896.11 in Table 1 and the AUC in Figure 5 decreases slightly by 0.001%. However, these differences are small, and additional data are needed to assess the significance of these differences.

Our models also provide some initial understanding of how the flash patterns are integrated over between spikes. Our PRO and PROs models allow the history dependence to be modeled in different ways, by changing the definition of $t^\ddagger$ and $t^*$ in Equation (2), and subsequently the definitions of CF and SF (PF only depends on the previous flash time). Then we can fit the fixed window PRO and PROs models using the newly defined predictors. To check how different ways of modeling the history dependence perform, we try two fixed time windows, by setting $t^\ddagger = t^* = t - \tau$, where the window size $\tau = 10$ bins (or 50 ms) or $\tau = 100$ (500 ms). For both choices, the residual deviance of the resulting models increases to 1086.7 and 901.66 receptively, compared with 895.38 in Table 1. The AUC values also drop to 87.9% (95% CI: 86.0%-89.8%) and 87.3% (95% CI: 85.5%-89.1%), compared with 92.7% in Figure 5. All the coefficients are significant when $\tau = 10$, and only the PF coefficient is significant (p < 0.05) when $\tau = 100$. Though the degeneration of model fit and prediction is small comparing with our simple choice, additional data may be required to optimize for the best choice of history dependence, which is beyond the scope of this paper. We also believe that the history dependence may depend on the experiment and neurons, and it should be studied on a case-by-case basis.

## 5. Simulations

In this section, we use simulated data to illustrate how well the PRO models can capture neural behaviors and predict future spikes. We want to assess the generalizability of the PRO functions described before, using simulated neurons from two models: Leaky Integrate-Fire (LIF). see [30] for example; and simple neuron network model (Network) [41]. The PRO approaches are compared with RF, SVM, GIF, and GLF. The emphasis of comparison will be generalizability, biophysical interpretation and computation speed.

5.1. *Simulations from Leaky Integrate-Fire*

The Leaky Integrate-Fire (LIF) model is a canonical model in cellular neuroscience. It employs the following differential equation to model the membrane potential, and assumes spikes occur whenever this potential up-crosses a fixed threshold. The membrane potential $V(t)$ follows

$$C\frac{dV(t)}{dt} + \frac{V(t)}{R} = I(t) \qquad (7)$$

where $I(t)$ is the stimulus processes, $C$ and $R$ are model parameters representing the membrane capacitance and resistance respectively. The resulting spike processes are given by putting a spike at the first time that $V(t)$ reaches a pre-set threshold $V_{th}$, and then $V(t)$ is reset to a small constant $V_{reset}$ immediately after the spike. We take the choice $V_{th} = 1$ and $V_{reset} = 0$. The time $t$ at which the threshold is reached is also recorded. $I(t)$ is set to be a box-shape stimulus (height=1, without loss generality), representing the flash pulses in the experiment. To simulate data from the dynamic model (7), we use 1/100 of our sampling rate (5ms) as step size, in a first-order method for Equation (7).

Each simulated dataset is generated for a period of 25s, at the same sampling rate of 5ms, which has the same dimension of our real experiment. That is, each dataset contains two sequences of 5,000 time points, one for the flashes and the other for the spikes. The same length input sequence $I(t)$ is randomly generated from iid Bernoulli(0.14), which indicate the 5ms bins when the box-shape stimulus is on. We set $C = 7$ and $R = 3$, because these values yield spiking sequences with similar summary statistics as the real experiment. These choices correspond to a membrane time constant of 21 msec.

Because PRO can be efficiently computed, we first fit the model to 10,000 simulation runs. Table 2 shows the frequencies of significant (at level 0.05) coefficients in PRO. The mean deviance R-squares of PRO fitting is 0.5681, with a standard error 0.0002. These results confirm that PRO generalizes to capture the spiking patterns from the classical LIF model. All the coefficients except for PF are significant for almost all the runs, where PF is significant over 80% of the runs. This may indicate that there is limited applicability of LIF to the optogenetics data, as this coefficient is highly significant in the real dataset. To summarize, PRO has good power of extracting non-zero coefficients, under the simulation model.

The intuition of how the PRO functions capture the integration of LIF is shown in Figure 6. We simulate the LIF model using two example flash sequences, starting from the same initializing point at time 0. The two sequences both have exactly two flashes after time 0, but one has a cluster of two flashes in two 5ms bins in [90, 100] ms while the other is evenly spaced at [45, 50] and [95, 100]. The SF function for the clustered flashes is consistently higher between [50, 100] with the conditional maximum at 100 ms (as also proved by Lemma 1), and the simulated membrane potential is also higher around 100 ms under this flash sequence, resulting in a spike. Due to the leaky property of LIF, the potential is smaller at 100 ms under two evenly distributed flashes and thus no spikes occur. Under this sequence, SF is also smaller, though the PF and CF values at time 100 are the same under both flash sequences. The PRO model can capture additional information not illustrated in this example, because the interaction term and the linear combination are used. The nonlinearity of LIF is further captured by the spline terms in PROs.

To gain scientific insights of these fitted coefficients, we replicate the previous simulation 100 times with the varying $C$ or $R$ scenarios. Because large changes of $C$ or $R$ may cause no firing at all during the whole observation period, we consider 11 equal-spaced relative changes from 80% to 120% of $C$ and $R$ respectively. In each scenario, we plot the fitted PRO coefficients of PF, CF, SF, and CF×SF respectively against the changing parameter ($C$ or $R$). A small fraction of insignificant coefficients are dropped from the plots, because PRO is not guaranteed to have 100% power in all scenarios, especially for the PF coefficient as we show in Table 2. The dropped percentages of coefficients are 9.27% PF in the varying $C$ scenario, 8.00% PF and 0.09% SF in the varying $R$ scenario.

Figure 7 plots the fitted PRO coefficients under varying $C$ and $R$ respectively. Note that the SF coefficients are smaller in the LIF simulations than the real experiment, and the PF coefficients are larger in magnitude. It is well known that the LIF neurons have a simple integration form given by Equation (7). Because PF characterizes short time effects and SF characterizes long time effects, the observed changes are consistent with the hypothesis that the neuronal circuits under optogenetic stimulation integrate the inputs in a sophisticated manner over a long period of time. Moreover, there may be some (possibly nonlinear) relations between the LIF parameters and the PRO coefficients. For example, there are deviations of the PF coefficients from the linear trend, and some deviations can be large in Figure 7. For simplicity, we employ simple linear regression to illustrate the overall linear relationship between these LIF parameters and the PRO coefficients. The trends with varying $C$ and $R$ (Figure 7) is clear except for PF, though all the slopes are significant at level $5 \times 10^{-5}$ or less. These trends may reveal important scientific insights. For example, when $R$ increases, decreasing the voltage leak, and thus the rate of increasing probability (due to increasing CF, or more flashes) increases, and the rate of increasing probability (due to increasing SF, or more closely paced flashes) increases.

We also compare the spike-to-spike prediction performance using the same LIF model. In each replication, we simulate 10,000 time points of flashes and spikes, where the first 5,000 points are used as the training dataset and the remaining 5,000 points are used as the testing dataset. Similar to the analysis of the real dataset, we compare the out-of-sample prediction performance on the testing data measured by AUC. Table 3 shows the average (SE) AUC values of different methods over 100 runs. Both PRO and PROs outperform all other methods, having the same best prediction performance and the smallest SE. The second best method is GIF but with the second largest SE.

Table 4 lists the computation times needed for model fitting and prediction of all methods. The computation is carried out on a machine with CPU Intel Core 2 6600 at 2.4GHz and 8 Gb of memory. PRO is particularly advantageous PRO is particularly advantageous as it takes only a small fraction of the computation time of all other methods. RF and SVM are second best, with almost the same computation time. PROs is the fourth fastest, and GLF and GIF are the slowest.

### 5.2. Simulations from Simple Neuron Networks

We also test our PRO models on a simple model of spiking neural networks [41], which enables fast simulations from a large network of neurons. We use the MATLAB code published in the paper, and we take the same set of network parameters, except for a few changes. All these changes are made in order to make the observed spiking behavior as close as possible to our real experiment. The purpose of this simulation study is not to provide a model for our optogenetic experiment, but rather to test the generalizability of our models under a neural network setting that may have similar spiking patterns as our experiment.

The network model from [41] is a two-dimensional ordinary differential equation model, and the differential equation is motivated by the Hodgkin-Huxley-type neuronal models. The model for a single neuron contains the membrane potential variable $V$ and the membrane recovery variable $U$. These two variables are related by the following differential equation

$$\frac{dV(t)}{dt} = 0.04V(t)^2 + 5V(t) + 140 - U(t) + I(t)$$
$$\frac{dU(t)}{dt} = a(bV(t) - U(t))$$

where the parameters including $a$ and $b$ are specified in the paper by fitting existing spike data on different types of neurons. Spikes are generated by thresholding $V(t)$, and both $V(t)$ and $U(t)$ are reset after spiking. This differential equation model is generalized to neural network as follows. The synaptic connections between neurons in a network is determined by a matrix $\boldsymbol{S}$. $S_{ij} = 0$ means that there is no connection from neuron $j$ to neuron $i$, and nonzero $S_{ij}$ measures the extent of synaptic connection between these two neurons. At time $t$, the input current $I_i(t)$ for neuron $i$ includes the stimulus input $f_i(t)$ and the total synaptic input, defined as

$$I_i(t) = \sum_{j:\text{neuron } j \text{ fires at time } t} S_{ij} + f_i(t).$$

The summation in the above equation models how neuron $i$ receives synaptic input from other neurons at each time point. The full details of this model are described in [41].

We simulate single-neuron spikes under network stimulation as follows. The network model consists of 1000 randomly connected neurons, where 800 are excitatory neurons and 200 are inhibitory neurons, and the synaptic weights $S_{ij}$ are independently set to either $0.5u$ if the $j$th neuron is excitatory or $-u$ if it is inhibitory, where $u$ is a randomly generated number from a uniform distribution between $[0, 1]$. To reduce the variation of the characteristics of excitatory neurons across simulation repeats, we choose the model parameters for the regular spiking type instead of the random type in the original paper. The positive feedback of each neuron is also removed to reduce clustering of spikes because no two spikes occurred within 5 ms in our experiment. Excitatory neurons are selected to be stimulated with probability 50%, and the spiking output is from a non-directly stimulated excitatory neuron that is the closest to the real experiment spiking frequency (350 spikes over 10s) per every 5 ms bin. If there are ties in spiking frequencies, we pick the one with the smallest number of spikes if the time bin is decreased to 1ms, or a random one if there are still ties after the decrease. During each randomly selected 5ms interval (with probability 0.14), each stimulated neuron is stimulated with independent Gaussians stimuli with mean zero and standard deviation 250, where the standard deviation is also picked in order to produce the closest spiking frequency. Outside the randomly selected 5ms intervals, the input is set to zero. From our simulations, these modifications lead to a similar spiking frequency as our experiment.

Table 3 shows that PROs predicts the best. GIF is the second best, followed by GLF and PRO. All these methods have AUC values smaller than 90% probably because the sampling neuron is randomly connected and the input is random as well. This simulation example tests the performance for possibly different neurons and neuronal networks with different properties. In all these cases, PRO has a decreased AUC probably due to increased non-linearity induced by the simulation model, but still outperforms black-box methods like RF and SVM. The computation time of these methods are compared in Table 4, where PRO is again the fastest one among all simulation scenarios. PROs, RF, and SVM have similar computational time, while GIF and GLF are thousands of times slower.

## 6. Discussion

This paper introduces a simple probabilistic model for optogenetics data. The PRO approach draws on successes of existing GLM framework for modeling spike train data. The innovation in the PRO model is to introduce explicit point-process responses to address challenges in modeling sparse input processes. The PRO model is shown to have good generalizability, interpretability and computation time. A spline extension, PROs, is also discussed, and is shown to have a similar performance.

Our experiment and model provide evidence that neurons in the recurrent layer V circuit in the prefrontal cortex integrate their inputs, i.e., a light flash is more likely to elicit a spike if there have been more previous light flashes since the last spike; see Figure 3. This suggests that the neurons are integrating their inputs in a sophisticated manner. The detailed mechanisms are of great interest but beyond the scope of our statistical model. A possible hypothesis is that each light flash does not just excite a neuron – it actually recruits activity in excitatory and inhibitory neurons, opening excitatory and inhibitory ion channels on the neuron. This then may further change its membrane time constant depending on the flash patterns. The LIF model clearly lacks such a mechanism, which may explain the reasons of slightly different PRO coefficients (e.g. smaller SF) despite it has the same spiking frequency. Our experiment and model provides initial evidence that the integration depends on the patterns of optogenetic input, where a light flash tends to enhance the chance that a subsequent light flash will elicit a spike. This leads us to hypothesize that the neuron must have mechanisms for remembering its history of input besides just the typical responses to excitatory and inhibitory input, which could be an interesting direction of future research. Moreover, the length of history dependence is also an important question, and our model is probably only providing a first order approximation on this issue. Further data are needed to check whether the dependence goes beyond the previous spike. We are currently conducting experiments to discover these biophysical mechanisms.

An interesting direction is to test the decoding capability of our model. We can simply modify our probability framework to capture the flash probability at each time given the history of spikes and flashes. This can be done by simply switching the notation $s(t)$ and $f(t)$ in the general formulation (1) and related equations afterwards. Similar considerations, such as reverse regression [24], exist for spike train data. It is also interesting to study the Bayesian decoding methods based on an GLM encoding [42].

It is interesting to further develop the PRO models for multi-neuron recordings. For computational efficiency, one can consider the conditional approach, which models how each neuron in a network responds to multiple point processes: the flash stimulus and the neural spiking of other recorded neurons. This approach is briefly described as follows. For neuron $j$, one can construct new response functions $PF_j$, $CF_j$, and $SF_j$ by replacing the flash process $\boldsymbol{f}$ with the spiking process $\boldsymbol{s}_j$ in Equation (4). For each recorded neuron $i$, one can expand our model to include these addition predictors $PF_j$, $CF_j$, and $SF_j$. The corresponding coefficients then have a natural interpretation in terms of how neuron $i$ integrates the synaptic inputs from other neurons in a network. All the model coefficients can be fitted over the sum of the conditional likelihood functions of all the neurons. This conditional approach is advantageous in computation, compared with modeling the distribution of large dimensional point processes. This approach has been employed before, see for example [15] and [41]. However, we note that there are experimental limitations to performing experiments to study multiple neurons in neural networks under optogenetic stimulation, e.g., photoelectric artifacts and synchronized firing; see [11] for a review of challenges. Once these issues are solved, it will be interesting to use models to study various characteristics of neural circuits, including those summary statistics provided by our models.

We are also interested in exploring additional biophysical meaning of the PRO coefficients. One interesting direction is to include some feedback component in the model. As we see from the PROs fit, the PF coefficient may deviate from the linear form, if a flash and spike occur within the same period. Additional predictors capturing this can be included to address this issue. It would be interesting to explore their biophysical meanings as well.

It would also be interesting to develop state space models for predicting the spike times, for example see [43, 18] for recent advances, and it would be interesting to incorporate these approaches in our models. Behseta and colleagues [44] also provide an important hierarchical model for spike train data, and it would be very interesting to extend that work for our optogenetics data.

## 7. Proof

### 7.1. Proof of Lemma *1*

*Proof.* Denote a fixed number $K = |F_t| > 2$. Let the general intra-flash time be positive integers $a_j = t_{(j-1)} - t_{(j)} > 0$ for $j = 2, \ldots, K$, because no two flashes are given in the same time. Additionally, the last one $a_K = t_{(K-1)} - t_{(K)} = t_{(K-1)} - t^\ddagger > t^* - t^\ddagger \geq 0$ because $t_{(K-1)} > t^*$ by the definition of $t^\ddagger$. The first one is the time past the last flash $a_1 = t - t_{(1)} \geq 0$ because it is possible $t_{(1)} = t_{(0)} = t$. For SF, it is equivalent to finding the maximum and minimum over $a_j$'s for the following objective function

$$L(a) = \sum_{j=2}^{|F_t|} \left(t_{(j-1)} - t_{(j)}\right)^2 + a_1^2 = \sum_{j=2}^{K} a_j^2 + a_1^2 \qquad (8)$$

under a fixed total sum constraint $M = \sum_{j=1}^{K} a_j = t - t^\ddagger$ and the constraints on $a_j$ stated before.

To derive the maximum of the objective function (8), let $a'_j = a_j - 1 \geq 0$ for $j = 2, \ldots, K$, and $a'_1 = a_1 \geq 0$. Then $\sum_{j=1}^{K} a'_j = M - K + 1$. Maximizing $L(a)$ is then equivalent to maximizing $\sum_{j=2}^{K} \left(a'_j\right)^2 + 2\sum_{j=2}^{K} \left(a'_j\right) + (a'_1)^2 = \sum_{j=2}^{K} \left(a'_j\right)^2 + 2(M - K + 1 - a'_1) + (a'_1)^2$. The positivity of $a'_j$ implies that

$$\sum_{j=2}^{K} \left(a'_j\right)^2 \leq \left(\sum_{j=2}^{K} a'_j\right)\left(\sum_{j=2}^{K} a'_j\right) \leq (M - K + 1)^2.$$

It is straightforward to see that the equality is achieved when one and only one $a'_j$ equals to $M - K + 1$ for any $j = 2, \ldots, K$, and the rest of $a'_j$ all equal to 0. This is then the configuration of $t_{(j)}$'s as stated.

Deriving the minimum of the objective function (8) is similar. The Hölder's inequality yields the following inequality

$$\sum_{j=1}^{K} a_j^2 \geq \frac{1}{K}\left(\sum_{j=1}^{K} a_j\right)^2 = \frac{M^2}{K},$$

where the lower bound is achieved under the solution

$$a_j^* = M/K \geq 1, \qquad (9)$$

provided that all the constraints of $a_j$ are also satisfied. To obtain an integer solution, we seek the integer grid points that are the closest to $a^*$ under the Euclidean distance, whose coordinates are thus either $\lfloor M/K \rfloor$ or $\lceil M/K \rceil$ satisfying the sum constraint $M$. □

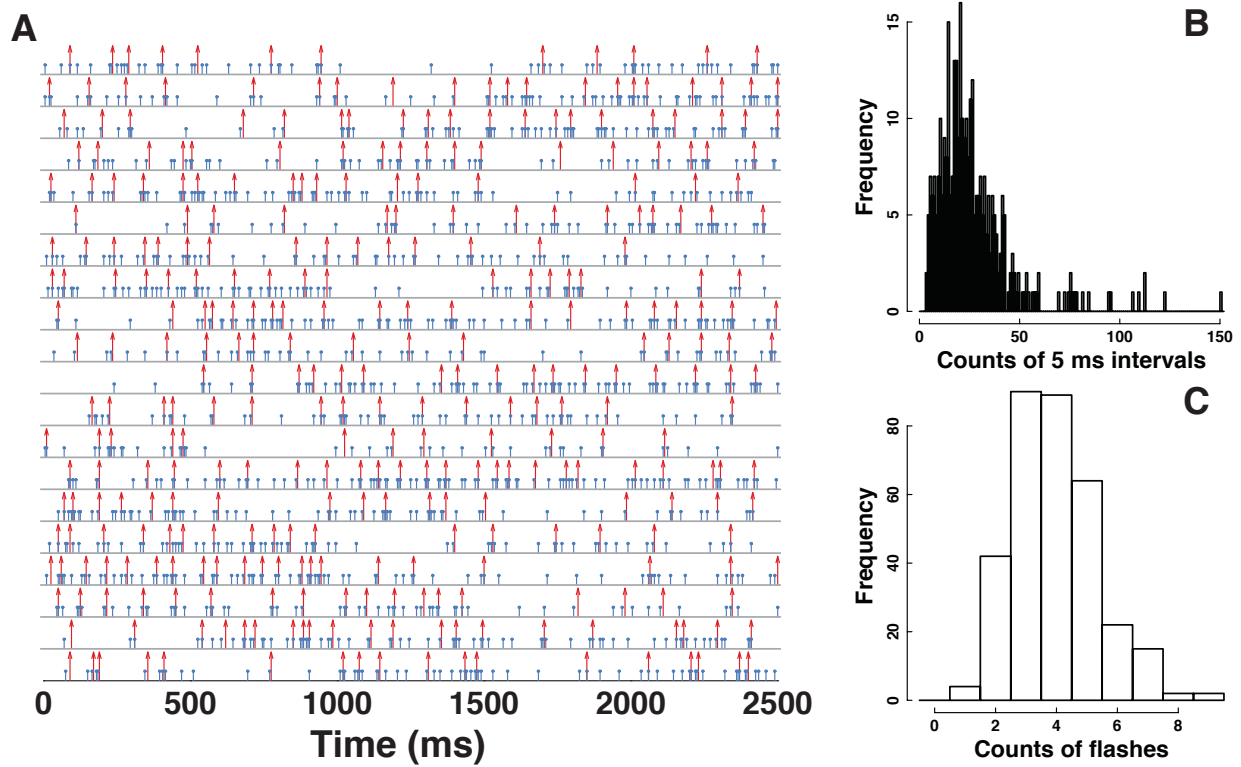

**Figure 1.** Spikes and flashes in an optogenetics experiment over 20 sweeps. Raster plot of spikes and flashes in 20 sweeps is in (A), above each horizontal line. Red long vertical arrows are spikes, and blue short vertical bars with solid circle heads are flashes. Histograms of inter-spike time (in counts of 5ms intervals) and total inter-spike flashes are in (B) and (C) respectively.

**Table 1.** Model fit and Wald standard errors (Z scores and p values) from PRO on the real experiment data. The null deviance is 1400.80 (df=4867), and the residual deviance is 895.38.

| Coefficients | Estimate | SE | Z value | P-value |
|---|---|---|---|---|
| Intercept | -24.310 | 4.115 | -5.907 | 3.47e-09 |
| PF | -1.447 | 0.134 | -10.796 | < 2e-16 |
| CF | 15.719 | 2.821 | 5.572 | 2.52e-08 |
| SF | 10.347 | 2.204 | 4.696 | 2.66e-06 |
| CF×SF | -7.108 | 1.507 | -4.716 | 2.41e-06 |

**Table 2.** Frequencies of significant PRO coefficients (at level 0.05) on the simulated LIF data. The standard error is not larger than 0.01 out of 10,000 replications.

| Coefficient for | PF | CF | SF | CF×SF |
|---|---|---|---|---|
| %(p-value<0.05) | 88% | 100% | 100% | 100% |

**Table 3.** Comparison of average (SE) AUC values (in percentages) over 100 simulations, from the LIF and Network models respectively. The best performance values are highlighted in bold. RF: random forests; SVM: support vector machine; GIF: generalized integrate-and-fire model; GLF: generalized linear model using filter bases.

| Model | PRO | PROs | RF | SVM | GIF | GLF |
|---|---|---|---|---|---|---|
| LIF | **97.50**(0.02) | **97.41**(0.03) | 92.02(0.18) | 90.20(1.49) | 94.20(0.85) | 81.09(0.19) |
| Network | 76.76(0.89) | **85.28**(0.76) | 71.62(0.86) | 68.34(2.25) | **82.06**(0.82) | 78.45(1.00) |

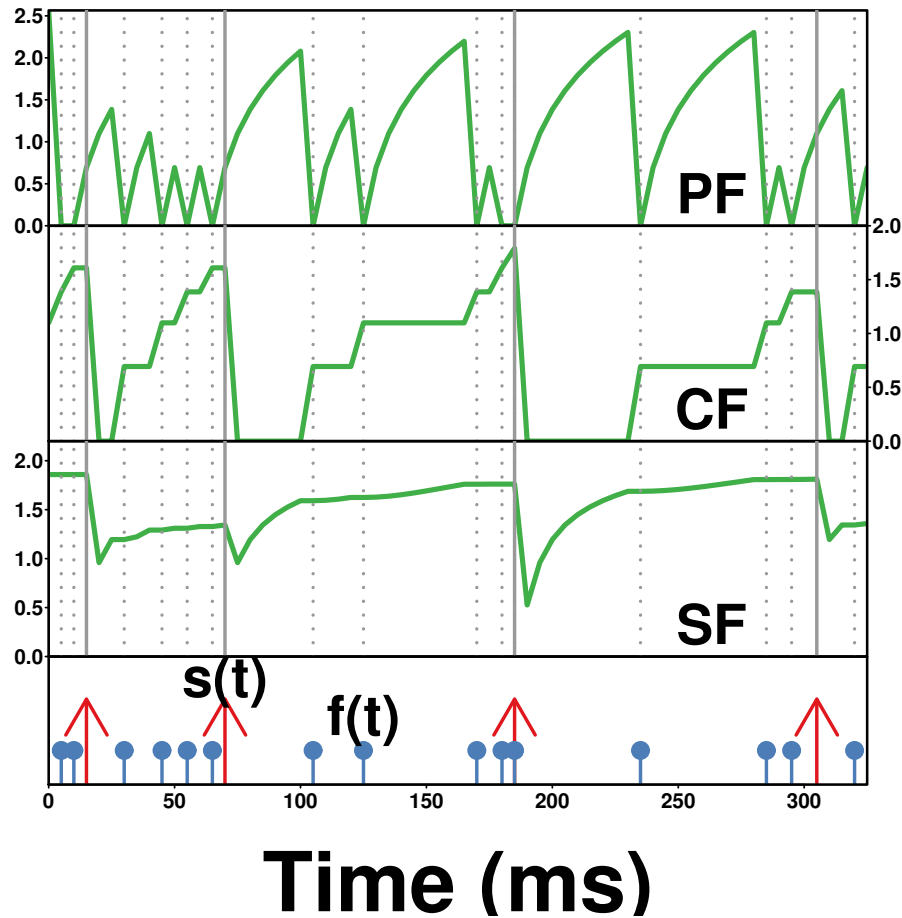

**Figure 2.** Schematic plots of 3 response functions (top 3 panels, in green) responding to an example sequence (bottom panel) of flashes $f(t)$ and spikes $s(t)$. Red long vertical arrows are spikes, blue short vertical bars with solid circle heads are flashes, grey dotted vertical lines represent the flash times, and grey solid vertical lines represent the spike times.

**Table 4.** Comparison of average (SE) computation times in seconds over 100 simulations, from the LIF and Network models respectively. The best performance are highlighted in bold. RF: random forests; SVM: support vector machine; GIF: generalized integrate-and-fire model; GLF: generalized linear model using filter bases.

| Model | Method | PRO | PROs | RF | SVM | GIF | GLF |
|---|---|---|---|---|---|---|---|
| LIF | Fitting | **0.195**(0.003) | 4.959(0.603) | 1.337(0.005) | 0.7416(0.005) | 469.80(5.49) | 181.75(13.41) |
| LIF | Prediction | **0.097**(0.0001) | 2.705(0.014) | 0.189(0.001) | 0.153(0.0004) | 3135.95(43.44) | 725.69 (5.83) |
| LIF | Total | **0.293**(0.003) | 7.664(0.604) | 1.527(0.007) | 0.894(0.005) | 3605.75(47.54) | 907.44 (15.91) |
| Network | Fitting | **0.153**(<0.001) | 3.119(0.079) | 1.592(0.01) | 0.975(0.02) | 500.37(7.89) | 725.116(33.10) |
| Network | Prediction | **0.091**(<0.001) | 0.729(0.002) | 0.258(0.001) | 0.169(0.001) | 3983.94(46.92) | 920.31(11.46) |
| Network | Total | **0.219**(<0.001) | 3.848(0.080) | 1.849(0.01) | 1.144(0.02) | 4484.31(51.62) | 1645.41(33.57) |

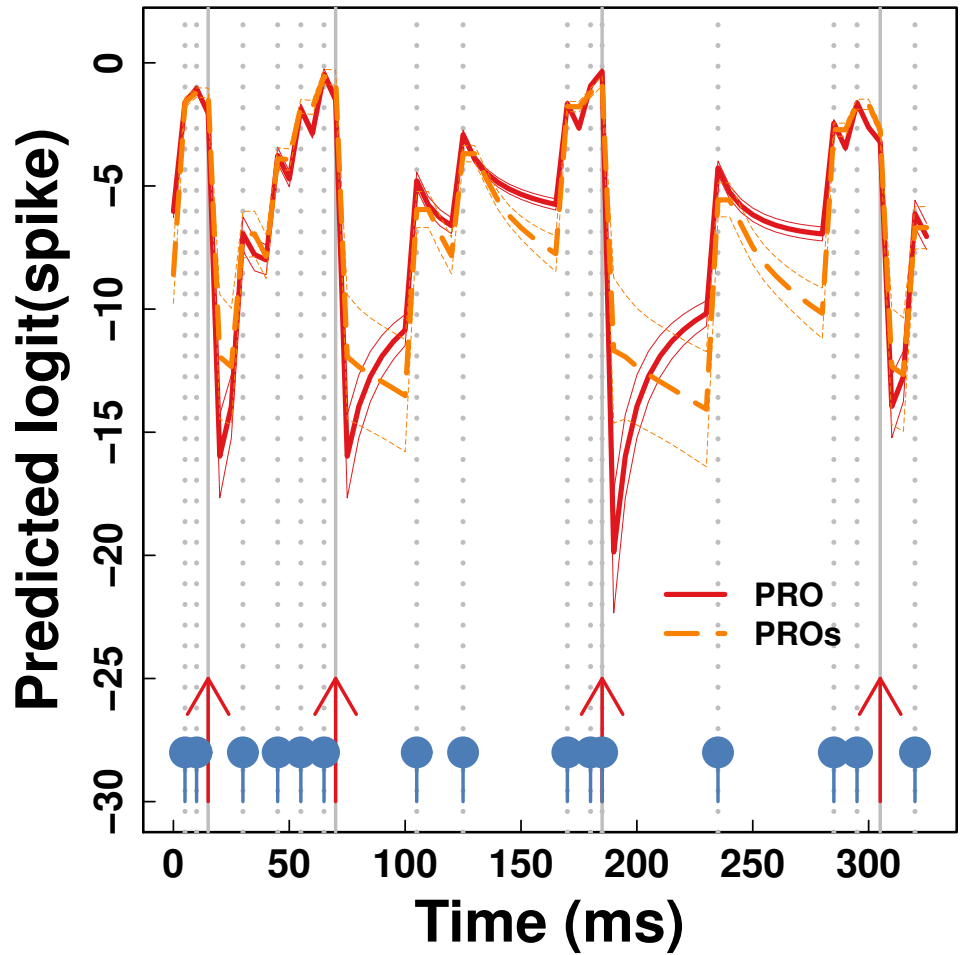

**Figure 3.** Predicted probabilities (logit) of spiking (bold lines) and the 95% confidence intervals (thin lines) from the PRO model (red, solid) in Table 1 and the PROs model (orange, dashed), responding to the example sequence (bottom) in Figure 2. Red long vertical arrows are spikes, blue short vertical bars with solid circle heads are flashes, grey dotted vertical lines represent the flash times, and grey solid vertical lines represent the spike times.

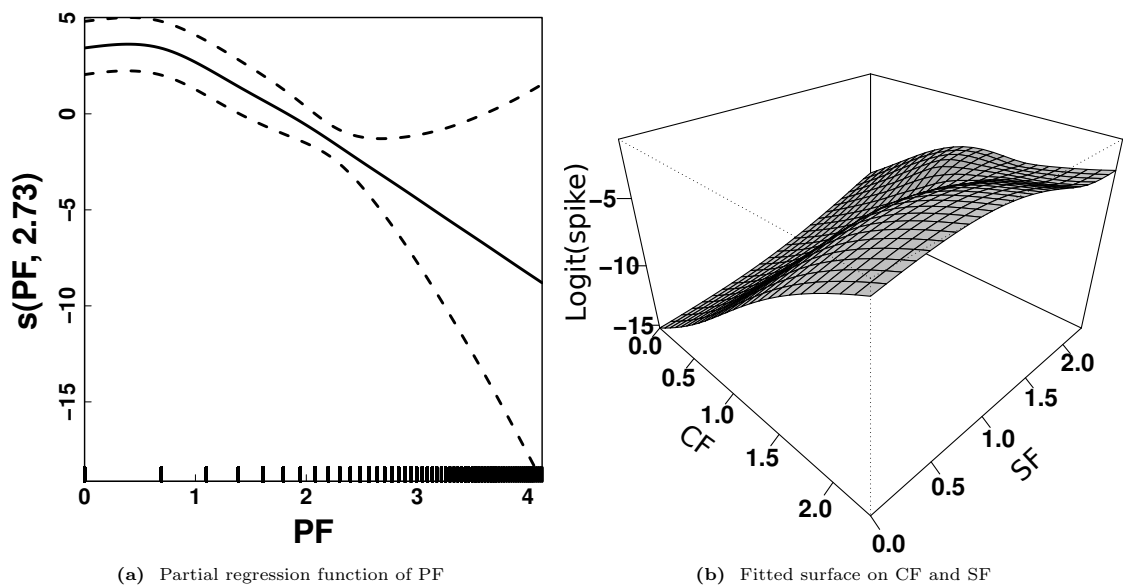

(a) Partial regression function of PF

(b) Fitted surface on CF and SF

**Figure 4.** Plots of the partial regression function of PF (left panel), and the fitted surface (in logits) on CF and SF with a fixed PF value at its mean (right panel). Both plots use only the first 10 sweeps of the real experiment data.

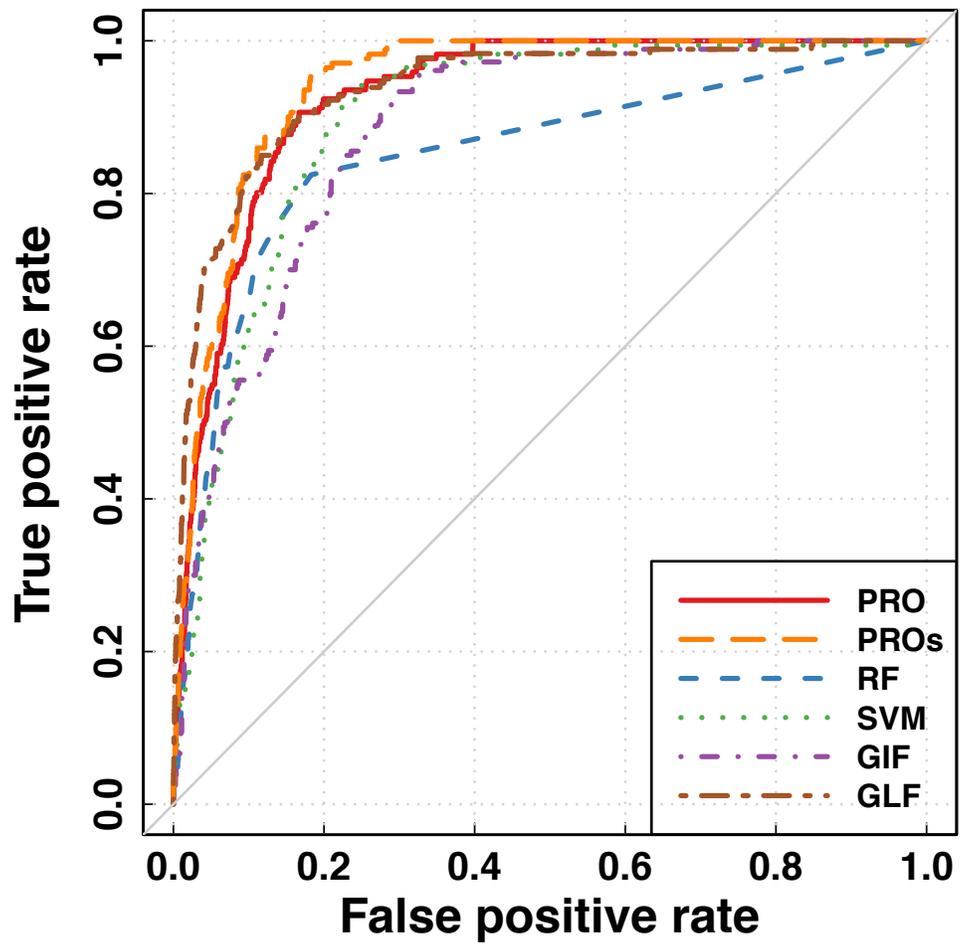

**Figure 5.** Plot of the ROC curves of all models. The AUC values are as follows: PRO=92.7%, PROs=94.2%, RF=85.0%, SVM=89.7%, GIF=88.1%, GLF=93.6%.

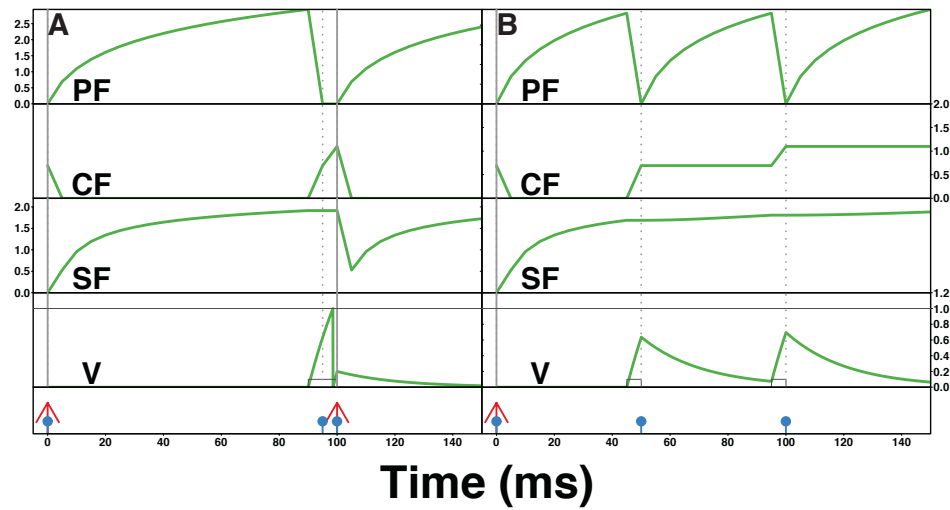

**Figure 6.** PRO functions and simulated membrane potentials (V) from the LIF model, under two example flash sequences: (A) two flashes between 90 to 100 ms; (B) one flash activated between 45 and 50 ms, and the other between 95 and 100 ms. SF is constantly larger in (A) than (B) between 50 and 100 ms. The plot of V is on the ms time scale while the PRO functions are on the 5ms time scale, and thus the flash inputs are shown as boxcar sequences in grey (scaled by 0.2).

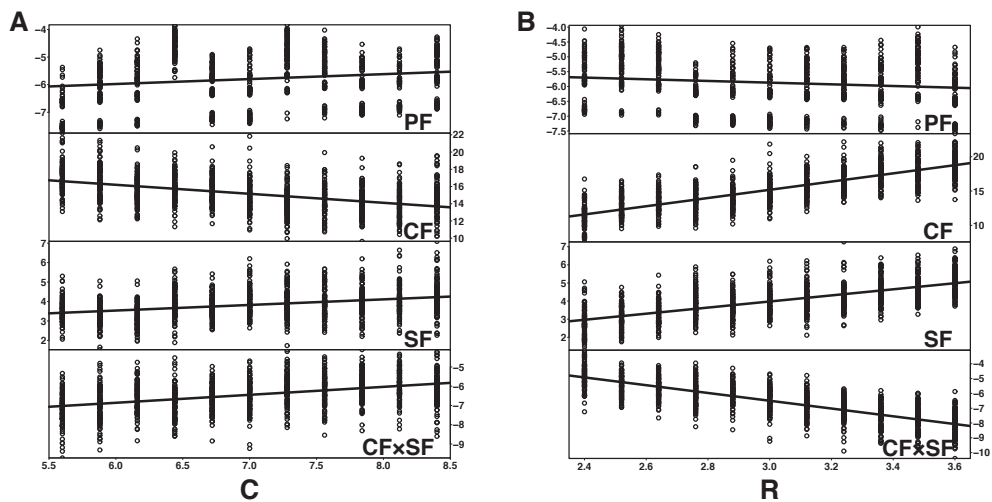

**Figure 7.** Plots of the PRO coefficients against the (A) varying $C$ and (B) varying $R$ parameter in the LIF model. (A) The linear regression slopes for each PRO function are 0.18 (PF), -1.04 (CF), 0.29 (SF), 0.41 (CF×SF), and all slopes are significant at level $10^{-6}$ or less. (B) The linear regression slopes for each PRO function are -0.28 (PF), 5.98 (CF), 1.68 (SF), -2.62 (CF×SF), and all slopes are significant at level $5 \times 10^{-5}$ or less.